\documentclass[preprint,aps]{revtex4}
\usepackage{amssymb,graphicx}
\newcommand{\be}{\begin{equation}}
\newcommand{\ee}{\end{equation}}
\newcommand{\bea}{\begin{eqnarray}}
\newcommand{\eea}{\end{eqnarray}}
\newcommand{\ba}{\begin{array}}
\newcommand{\ea}{\end{array}}

\newcommand{\eq}[1]{(\ref{#1})}
\begin{document}
\title{The influence of step-edge barriers on the morphological relaxation of
nanoscale ripples on crystal surfaces.}
\author{V.~B.~Shenoy, A.~Ramasubramaniam, H.~Ramanarayan, D.~T.~Tambe, W-L.~Chan and E.~Chason}
\affiliation{Division of Engineering, Brown University,
Providence, RI 02912}
\date{\small \today}

\begin{abstract}
 We show that the decay of sinusoidal ripples on crystal surfaces, where mass transport is limited by the attachment and detachment of atoms at the step-edges, is
remarkably different from the decay behavior that has been
reported until now. Unlike the decreasing or at most constant rate of amplitude decay of sinusoidal profiles observed in earlier work, we find that the decay rate {\em increases} with decreasing
amplitude in this kinetic regime. The rate of shape invariant amplitude relaxation is shown to be
inversely proportional to both the square of the wavelength and the current amplitude. 
We have also carried out numerical simulations of the relaxation of realistic sputter ripples. 
\end{abstract}
\maketitle

Nanoscale semiconductor devices hold potential for a wide range of
applications in optics, electronics and information technology. The shape evolution of these structures at typical growth temperatures takes place through mass transport via surface diffusion. With increasing technological interest in smaller scale structures, a fundamental understanding of the kinetics of this phenomenon is important, since small feature sizes generally lead to more rapid evolution of the shape. Material parameters that govern atomic surface kinetics are efficiently determined from morphological relaxation of simple shapes, for example periodic sinusoidal surface ripples, which can be produced by lithographic pattering or are formed as a result of surface instabilities during growth, sputtering or etching. In the case of crystal surfaces above the roughening temperature and amorphous surfaces, the curvature based continuum theory developed by Herring and Mullins \cite{Herring} and the experiments supporting this theory for sinusoidal ripples, notably by Blakely and coworkers \cite{Blakely} are considered classic studies in surface science.

Surface evolution in the case of crystal surfaces below the roughening temperature takes place through diffusion of atoms on terraces and by their attachment and detachment from the step-edges. The surface energy in this case is determined by the
formation and the interaction energies of steps and consequently
acquires a cusp at facet orientations. Due to the presence of the strong non-linearity
that arises from the cusp, extension of the
classical theory to faceted surfaces is non-trivial. Earlier work on sinusoidal ripples, both continuum studies \cite{Villain,Zangwill,Spohn} and kinetic Monte Carlo (KMC) simulations \cite{Murthy,Erlebacher} focus on the regime where the kinetics of mass transport is determined by terrace-diffusion limited (TDL) kinetics. In the case of step attachment-detachment limited (ADL) kinetics, attention has been restricted to the relaxation of 1D sinusoidal profiles \cite{Cahill,Kandel}, where the line-tension of steps does not appear in the analysis. Many systems considered in recent experiments, notably 2D ripples on Si(001) \cite{Erlebacher1}, Cu(001)\cite{Chason}, Au(001) and Ni(001) \cite{Blakely}, where the line-tension contribution should be significant due to the curvature of steps, are believed to be in the ADL regime. 

In this article, we consider the relaxation of 2D sinusoidal ripples (symmetric and elongated) and show that the decay rate of the
amplitude {\em increases} with decreasing amplitude in the ADL case, for non-zero values of the formation energy (or line-tension) of steps. This is not an immediately obvious result, but can be
established
by numerical solution of the continuum surface evolution equations, scaling analysis and KMC simulations. We also carry out simulations of relaxation of realistic sputter ripples that consists of several Fourier modes in the vicinity of a dominant peak in the surface spectrum. 

We start our continuum approach by relating the surface mass flux on the stepped surface to the gradient of the surface chemical potential $\mu$, using the linear kinetic relation ${\mathbf j}=-C(\nabla h)~\nabla \mu$, where $h$ denotes the height of the surface measured from a nominally flat reference plane, $\nabla$ denotes the surface gradient and $C$ is the surface mobility parameter. The mobility parameter is taken to be of the form
\begin{equation}
C(\nabla h)=\frac{D_sC_{eq}}{k_BT}\left(1+\frac{2D_s|\nabla h|}{\kappa h_s}\right)^{-1},
\label{mobility}	
\end{equation} 
in which $D_s$ is the diffusion constant of adatoms, $\kappa$ is the effective rate of attachment and detachment of atoms at a step-edge, $h_s$ denotes the height of steps and $C_{eq}$ is the equilibrium adatom concentration ahead of a straight non-interacting step at temperature $T$. This functional form is obtained by taking the continuum limit of the terrace mass flux from the classic Burton-Cabrera-Frank model for discrete surface steps \cite{BCF,Zangwill}. The relative importance of the terrace diffusion and the attachment-detachment processes can be obtained by comparing the effective length parameter $l=D_s/\kappa$ with the terrace width, $\Delta x = h_s/|\nabla h|$.
For $l \ll \Delta x$, the kinetics of mass transport is diffusion
limited, whereas in the opposite limit, $l\gg \Delta x$, ADL kinetics
is obtained. In the former case, the
mobility parameter is simply a constant, $D= \frac{D_sC_{eq}}{K_BT}$, while in the latter case, the mobility parameter is inversely proportional to the
local slope. This recovers
the result first derived by Nozi\`{e}res~\cite{Nozieres} for ADL kinetics. The expression in \eq{mobility} describes a smooth transition between these two limiting cases.

The driving force for surface diffusion can be obtained from the surface chemical potential, which can be written as 
\begin{equation}
\mu({\bf x})=-\frac{\partial}{\partial x_i} \bigg(\frac {\partial\gamma(\nabla h)}
{\partial h_{x_i}}\bigg),
\end{equation} 
where $\gamma(\nabla h)=\gamma_0+\beta_1|\nabla h|+{\beta_3}|\nabla h|^3$ is the usual form for the surface energy of a stepped surface in the small-slope limit. Here, $\gamma_0$ is the surface energy of the facet, $\beta_1$ is related to the formation energy per unit length of a step and $\beta_3$ is related to the interaction between two steps. If the surface chemical potential is known, conservation of mass gives the equation for the evolution of surface height,
\begin{equation}
\frac{\partial h}{\partial t} = -\nabla \cdot \left[C(\nabla h) \nabla \mu({\bf x})\right].
\label{evoleqn}
\end{equation}
Because of the presence of a cusp in the surface energy, the chemical potential becomes singular as $|\nabla h| \rightarrow 0$, which leads to difficulties in the numerical solution of \eq{evoleqn} using standard discretization methods, unless the singularity is regularized. The solution, however can be accomplished without a regularization of the surface energy using a recent variational formulation of surface evolution \cite{Shenoy1}. The basic idea
of the formulation is to express the surface shape in terms of a Fourier series expansion and to use a variational principle to obtain coupled nonlinear ordinary differential equations (ODEs)
for the expansion coefficients. If the initial surface shape
is known, the evolution of these coefficients can be obtained using standard numerical integration
techniques for ODEs. 
The convergence of the solutions can be assessed by varying the number of terms in the Fourier expansion. 
Below, we will analyze the relaxation of sinusoidal ripples in the case of ADL kinetics, first using the variational method, followed by a scaling analysis and KMC simulations.

\begin{figure}[h!]
\begin{center}
\includegraphics[width=0.8\textwidth]  {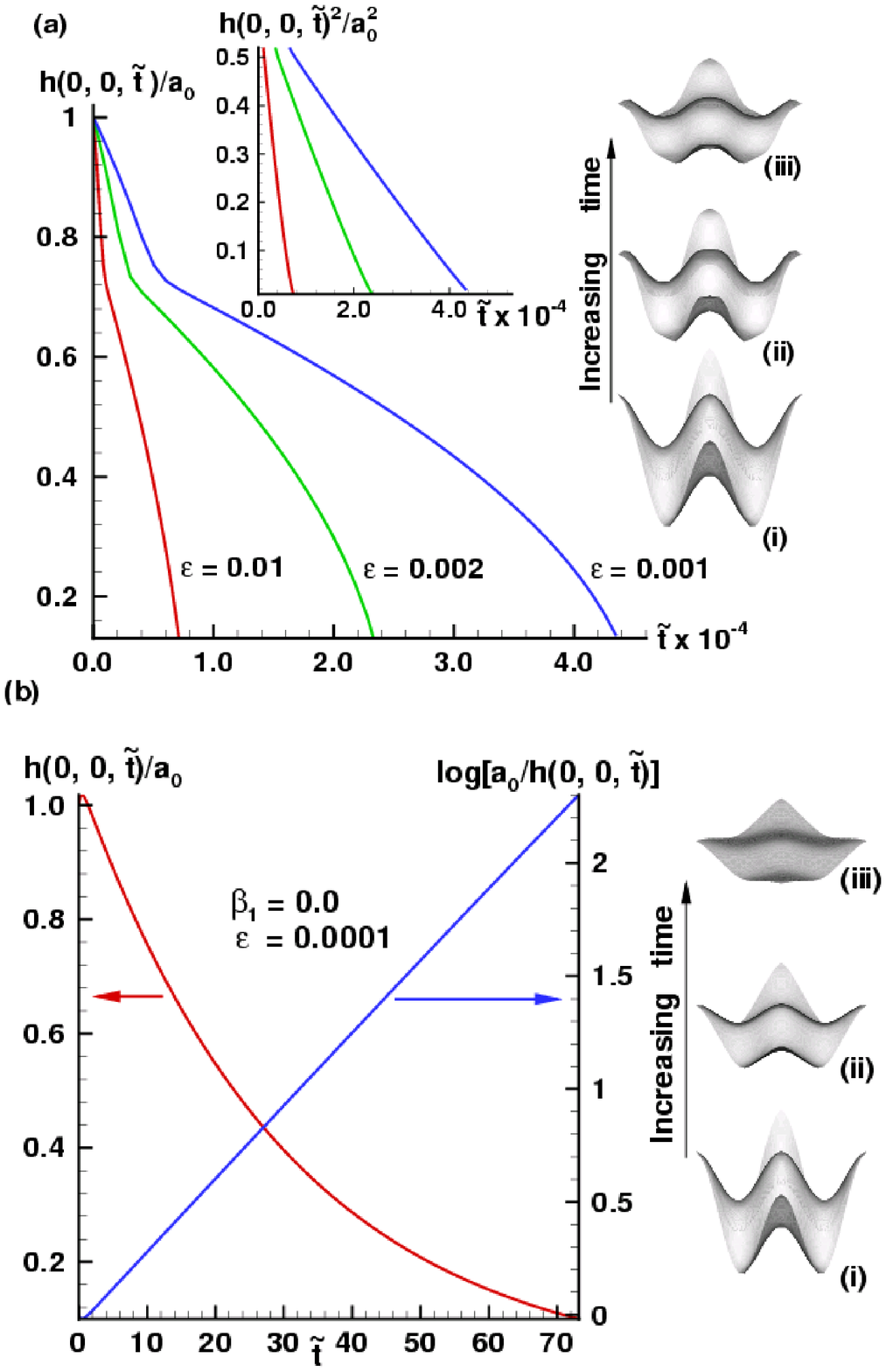}
\vspace{-1.5cm}
\caption{ Amplitude relaxation for 2D profile $h(x_1,x_2,t=0)=a_0\cos(kx_1)\cos(kx_2)$, where $k$ is the wavenumber and $a_0 = 0.01/k$. The relaxation is plotted as a function of $\tilde{t} = D\beta_3k^4t$, for $\tilde{\beta_1} = 1$ and $0$ in (a) and (b), respectively. As the
parameter $\epsilon = \kappa h_s/2D_s$ in (a) decreases, the relaxation is expected to follow ADL kinetics. The inset graph shows the same
curves, but with the vertical axis now set to $h^2(0,0,\tilde{t})/a_0^2$ in the region where the profile shows shape preserving relaxation -- the curves are seen to be linear for small $\epsilon$, in agreement with Eq.~\eq{scaling}$_1$. In contrast to (a), the decay in (b) is seen
to be exponential in time, as evidenced by the linear dependence of the logarithm of the amplitude on time. The time sequences show facet formation (insets (ii) and (iii)) in (a), while the profile sharpens at the extrema in (b). 
}
\label{fig1}
\end{center}
\vspace{0cm}
\end{figure}

  The influence of the step-edge barriers on surface relaxation can be conveniently discussed in terms of the dimensionless parameter $\epsilon = \kappa h_s/2D_s$, so that ADL kinetics would be observed as this parameter becomes small. The amplitude relaxation of a small amplitude 2D sinusoidal ripple is plotted in Fig.~1 for three different values of this parameter,  by choosing the ratio $\tilde{\beta_1} \equiv \beta_1/\beta_3=1$.
In all cases, we find that after an initial transient during which facets develop and grow at the extrema of the profile (the amplitude decreases by $\sim$ 20\% in this period as shown in Fig.~1(a)), the shape relaxes in a self-similar manner at a rate that {\em increases} as the amplitude decreases in magnitude \cite{Modes}. We have also studied amplitude relaxation for different values of the ratio $\tilde{\beta_1}$ and for sinusoidal profiles for which the ratios of the modulating wavelengths in the coordinate directions are different from unity. We find the decay behavior for finite values of $\tilde{\beta_1}$ to be fundamentally different from the decay for $\tilde{\beta_1} = 0$. In the latter case, 
because of the absence of energetic cost for formation of steps, 
the decay shapes do not show any facets, but on the contrary, the top of the profiles become sharper in time, as shown in Fig.~1 (b). While the amplitude relaxation in this case is exponential (refer to Fig.~1(b)), any non-zero value of $\tilde{\beta_1}$ leads to a behavior similar to Fig.~1(a).
These results indicate that the case $\tilde{\beta_1} \ne 0$ is a singular perturbation of the special case with $\tilde{\beta_1} = 0$. The observations regarding the functional form of relaxation do not change if the ratio of modulating wavelengths along the coordinate directions are changed - this result is similar to our earlier work in the TDL case \cite{Shenoy1}, where decay behavior was shown not to depend on the aspect ratio of the ripples. We also note that 1D sinusoidal profiles (like the 2D case in Fig.~1(b)) decay in an exponential manner when $\tilde{\beta_1}=0$, which is in agreement with earlier work \cite{Cahill,Kandel} where only the contribution from step interactions was included in the surface chemical potential.

Progress in understanding the surprising behavior in the case of ADL kinetics can be made using scaling arguments based on two key observations.  
First, for shape preserving relaxation, we can write the surface height as $h({\bf{\tilde{x}}},t) = a(t)\tilde{h}({\bf{\tilde{x}}})$, where ${\bf{\tilde{x}}} = k(x_1,x_2)$ denotes scaled coordinates and $k=2\pi/\lambda$ is the wavenumber of the modulation considered in Fig.~1. Next, for $\beta_1 \ne 0$, just as in the case of TDL kinetics \cite{Murthy,Villain}, our numerical simulations show that the decay rate is insensitive to the value of the interaction strength $\beta_3$ during shape invariant relaxation. If the corresponding term in the surface chemical potential is assumed to be small $a(t)$, and $\tilde{h}({\bf{\tilde{x}}})$ satisfy 
\begin{equation}
\dot{a}(t) = -\frac{k^2}{a(t)}\left(\frac{\kappa\beta_1h_sC_{eq}}{2k_BT}\right) \,\, {\mathrm {and}} \,\, \tilde{h} = \frac{\partial}{\partial{\tilde{x}_i}}\frac{1}{|\tilde{\nabla}\tilde{h}|}\frac{\partial^2}{\partial {\tilde{x}_i}\tilde{x}_j}\frac{\partial |\tilde{\nabla}\tilde{h}|}{\partial\tilde{h}_{\tilde{x}_j}},
\label{scaleq}
\end{equation}
respectively, where the partial differential equation for $\tilde{h}({\bf{\tilde{x}}})$ does not depend on any physical parameters and the solutions are to be found in the domain $0 \le \tilde{x_1},\tilde{x_2} \le 2\pi$. Since this equation is singular, it has to be solved by using its weak form or by employing variational methods introduced in  Ref.~\cite{Shenoy1}- the solution is very close to the faceted profile marked (iii) in Fig.~1(a). 
The ODE for $a(t)$ can be readily integrated to obtain 
\begin{equation}
a(t) = a(0)\sqrt{1-\frac{t}{\tau}}, \,\, {\mathrm{where}} \,\,
\tau = \frac{a(0)^2}{k^2} \left(\frac{2k_BT}{\kappa\beta_1h_sC_{eq}}\right).
\label{scaling}
\end{equation}
The above analysis shows that the square of the amplitude should decrease linearly in time during shape invariant decay, which indeed is borne out by the numerical calculations using the variational approach (refer to the inset in Fig.~1(a)). A physical argument for the increase in decay rate with amplitude can be provided by noting that the material from the top of the profile that has to make its way to fill in the bottoms during relaxation, encounters fewer steps at later stages of decay compared to the early or intermediate stages. Since each step presents a barrier to the atoms that are moving down \cite{Schwoebel}, an increase in decay rate should then be expected with decreasing amplitude, or, alternatively, the total number of steps between the tops and bottoms of the profile.

\begin{figure}[h!]
\hspace{0cm}
\begin{center}
\includegraphics[ width=1.1\textwidth] {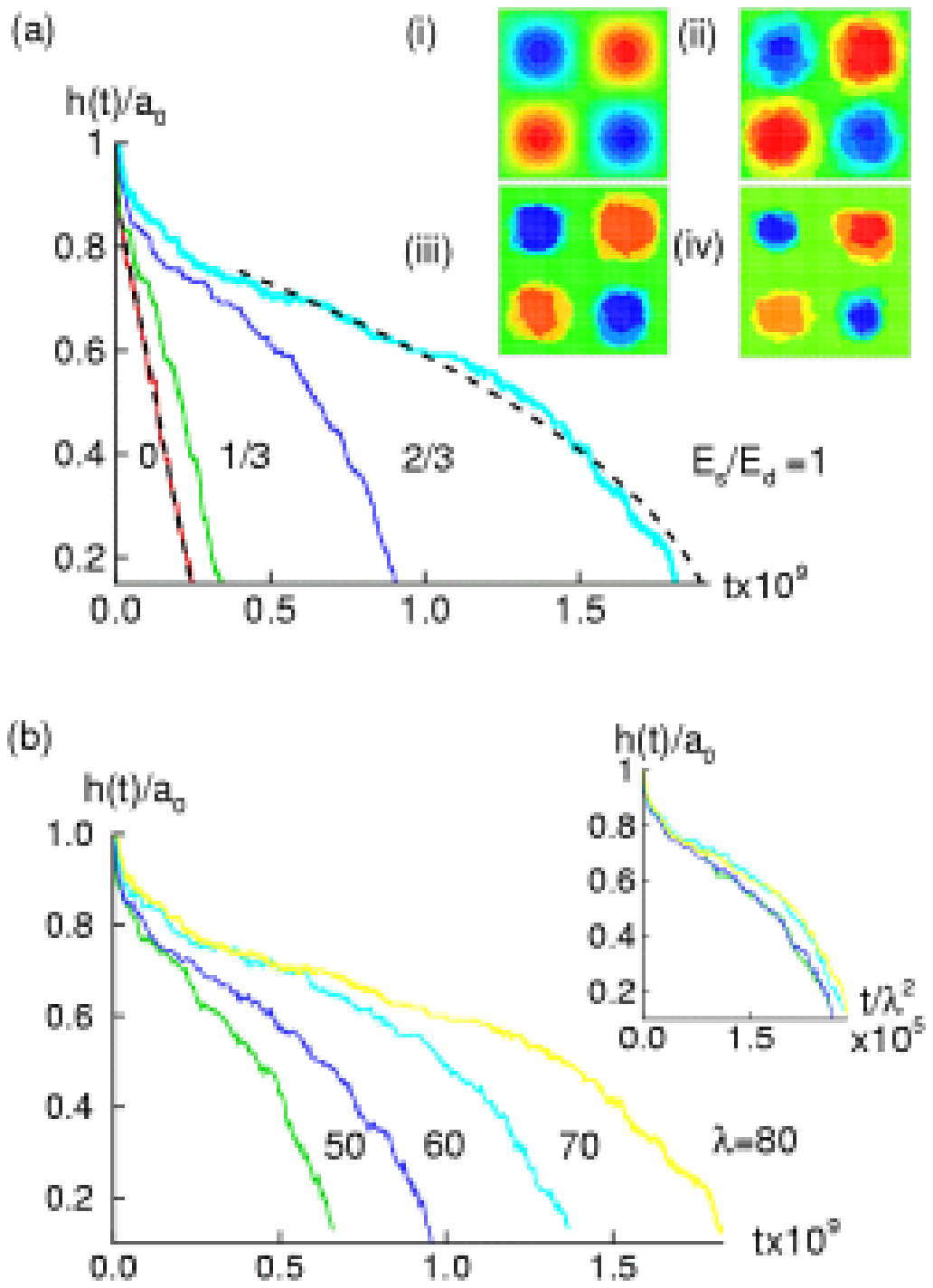}
\vspace{-10cm}
\caption{(a) The normalized amplitude $h(t)/a_0$ plotted as a function of KMC time
$t$ for different ratios of the step barrier ($E_s$) to the diffusion barrier ($E_d$) for a 2D sinusoidal profile with amplitude and wavelength of 6 and 80 lattice spacings, respectively. In this calculation, we have chosen $T = 0.41 T_R$ and set $E_d = \epsilon$, the bond strength in the SOS model. The dashed lines for  $E_s/E_d$ = 0 and 1 represent fits to linear decay for TDL kinetics \cite{Villain} and the decay law in Eq.~\eq{scaling}$_1$, respectively. 
 The insets show the evolution of surface morphology for $E_s/E_d=1$,
starting from the initial shape (i). 
(b) The normalized amplitude plotted as a function of KMC time
for different wavelengths $\lambda$ (in lattice units) for the case $E_s/E_d =1$ at $T=0.41 T_R$.
The inset shows that curves for different wavelengths nearly collapse to a single curve if time is scaled with the square of the wavelength.
}
\vspace{-0cm}
\label{fig2}
\end{center}
\end{figure}

We now turn to an analysis of ripple relaxation using KMC simulations with a simple cubic solid-on-solid(SOS) model with near neighbor interactions. Here, we start with a 2D sinusoidal ripple shown in the inset in Fig.~2(a) and allow the atoms to perform hops at a rate determined by the number of neighbors and the barriers that they encounter during an atomic jump. The jump rate for an atom can be written as $\nu_0 \exp[-(n\epsilon+E_b)/k_BT]$, where $\nu_0$ is an attempt frequency, $n$ is the number of bonds attached to the atom in consideration, $\epsilon$ is the bond strength and the barrier $E_b$ is equal to the diffusion barrier $E_d$ for hops on the same terrace, and is taken to be $E_d + E_s$ when an atom steps up or down a terrace, where $E_s$ is the additional barrier due to the Schwoebel effect \cite{Schwoebel}. Our calculations are similar to the work of Erlebacher and Aziz \cite{Erlebacher} who considered the case $E_s =0$, but here we study surface relaxation as the parameter $E_s$ is varied. Since the attachment-detachment rate $\kappa$ in Eq.~\eq{mobility} decreases in an exponential manner with 
$E_s$, the system would get closer to the ADL regime as this parameter increases in magnitude.

The result of a typical relaxation simulation is shown in the inset in Fig.~2(a) for a system with an initial amplitude ($a_0$) and wavelength of 6 and 80 lattice spacings, respectively. The profile shows a layer by layer decay, with very little step motion in other layers as the top and bottom layers are annihilated. The time dependence of the amplitude decay obtained by averaging twenty five statistically independent simulations at $T = 0.41 T_R$, where $T_R= 0.62 \epsilon$ is the roughening temperature, is shown in Fig.~2(a). When $E_s =0$, we recover the linear decay observed in earlier KMC simulations \cite{Erlebacher,Murthy}, while the rate of decay increases with decreasing amplitude for larger values of $E_s$. A fit in the case of $E_s/E_d=1$ shows that the amplitude decay is in agreement with the behavior predicted in Eq.~\eq{scaling}. Further evidence for the scaling given in Eq.~\eq{scaling} can be found in Fig.~2(b), where for a fixed amplitude \cite{Amplitude}, the relaxation time scales with the square of the wavelength in the ADL, in contrast to cubic scaling observed in the TDL case \cite{Villain,Murthy,Erlebacher}. 

\begin{figure}
\hspace{0cm}
\begin{center}
\includegraphics[ width=0.8\textwidth] {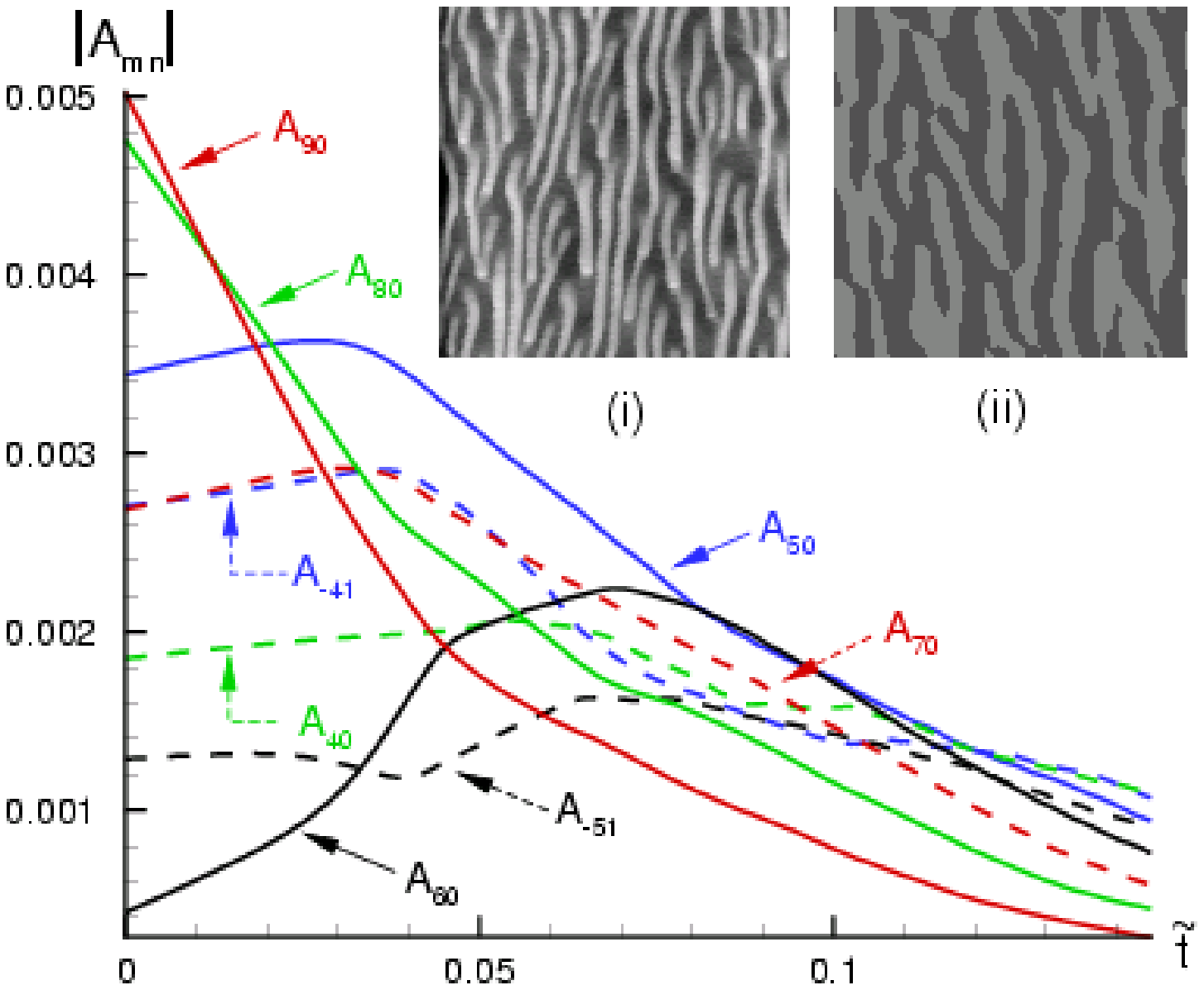}
\vspace{-8cm}
\caption{
Scaled magnitudes of dominant Fourier modes  $|A_{mn}| = k|\tilde{A}_{mn}|$ in the expansion $h(x_1,x_2,t) = \sum_{m,n}\tilde{A}_{m,n}\exp{ik(mx_1+nx_2)}$ for the shape of a rippled surface plotted as a function of rescaled time $\tilde{t} = D\beta_3k^4t$
for $\tilde{\beta}_1=1$ in the ADL regime. Inset (i) shows a typical AFM image of an unannealed
sputter-rippled Cu surface. The size of the image is 5$\mu$mx5$\mu$m with an
initial ripple wavelength and amplitude of $\approx$ 300 nm and $\approx$ 20 nm, respectively. Inset (ii) shows the unannealed sample of the same size used in the numerical study, with a dominant
wavelength of 500 nm and an amplitude of 25 nm. 
}
\vspace{-0cm}
\label{fig3}
\end{center}
\end{figure}

Finally, we consider the relaxation of realistic sputter ripples studied in recent experiments on Si and Cu surfaces \cite{Erlebacher1,Chason}. While these ripples are characterized by a dominant wavelength, there is a  spread in the surface roughness spectrum, due to the lack of perfect periodicity as shown in Fig.~3(i). Since the surface evolution equation \eq{evoleqn} is non-linear, a natural question to address is whether the decay of individual modes are influenced by the presence of neighboring modes. To this end, we have considered the evolution of the rippled surface in Fig.~3(ii) in the limit of ADL kinetics, by keeping track of approximately 400 modes using the variational approach \cite{Shenoy1}. As the surface relaxes, smaller wavelength modes decay more rapidly, while the modes with larger wavelengths progressively dominate the spectrum. While a dominant wavelength can be identified at any given instant of time, its evolution neither follows the form given in \eq{scaling} nor shows an increasing rate of decay with amplitude. It can also be seen that some of the long wavelength modes, initially grow in amplitude before they begin to decay. A similar trend has been recently observed on ripples on Cu surfaces \cite{Chason}. The above analysis shows that the non-linear coupling between modes on a typical rippled surface leads to complex relaxation behavior, and simple scaling rules for idealized sinusoidal corrugations cannot be directly applied to individual Fourier modes. We have also carried out similar analysis for TDL kinetics and found that the decay of dominant modes is not described by the linear relaxation behavior \cite{Villain,Murthy,Erlebacher} of a single sinusoidal mode. It is possible that other metrics, for example the integrated amplitude of the power spectrum over all modes or the evolution of the peak and width of the spectrum might show trends that could be modeled with simple functional fits. We hope to explore these directions in the future.

In conclusion, we have shown that the decay of sinusoidal profiles on crystal surfaces in the ADL limit, shows an increasing rate of relaxation with decreasing amplitude. This surprising behavior is attributed to kinetics of mass transport on stepped surfaces, in particular, the dependence of the surface mobility on the local density of steps. Analysis of rippled surfaces considered in experimental studies shows significant mode-coupling effects owing to the inherent non-linearity of the surface evolution
equations.

The research support of the National Science Foundation
through grants CMS-0093714 and CMS-0210095 and the Brown University
MRSEC program through grant DMR-0079964 is gratefully acknowledged.


\begin{thebibliography}{30}
\bibitem{Modes}
We have used 256 Fourier modes in this calculation, but almost identical results can be obtained with only 100 modes.
\bibitem{Amplitude}
We have also used KMC simulations to confirm that for fixed wavelengths, data collapse is obtained when time is scaled with the square of the amplitude (refer to Eq.~\eq{scaling}), as opposed to the amplitude in the TDL case \cite{Villain}.
\end{thebibliography}
\end{document}